\documentclass[%
 aip,
 jmp,
 amsmath,amssymb,
 reprint
]{revtex4-1}

\usepackage{graphicx}% Include figure files
\usepackage{dcolumn}% Align table columns on decimal point
\usepackage{bm}% bold math
\usepackage{color}
\usepackage{epstopdf}
\usepackage{amsmath}
\usepackage{mathtools,ulem}

\usepackage{natbib}

\begin{document}

\preprint{AIP/123-QED}

\title[AE stability in Multi NBI plasma]{Study of Alfven Eigenmodes stability in plasma with multiple NBI driven energetic particle species}

\author{J. Varela}
\email{jacobo.varela@nifs.ac.jp}
\affiliation{National Institute for Fusion Science, National Institute of Natural Science, Toki, 509-5292, Japan}
\affiliation{Oak Ridge National Laboratory, Oak Ridge, Tennessee 37831-8071}
\author{D. A. Spong}
\affiliation{Oak Ridge National Laboratory, Oak Ridge, Tennessee 37831-8071}
\author{L. Garcia}
\affiliation{Universidad Carlos III de Madrid, 28911 Leganes, Madrid, Spain}
\author{Y. Todo}
\affiliation{National Institute for Fusion Science, National Institute of Natural Science, Toki, 509-5292, Japan}
\author{J. Huang}
\affiliation{Institute of Plasma Physics, Chinese Academy of Science, Hefei, China}
\author{M. Murakami}
\address{Oak Ridge National Laboratory, Oak Ridge, Tennessee 37831-8071}

\date{\today}% It is always \today, today,

\begin{abstract}

The aim of this study is to analyze the destabilization of Alfven Eigenmodes (AE) by multiple energetic particles (EP) species in DIII-D and LHD discharges. We use the reduced MHD equations to describe the linear evolution of the poloidal flux and the toroidal component of the vorticity in a full 3D system, coupled with equations of density and parallel velocity moments for the energetic particles species, including the effect of the acoustic modes, diamagnetic currents and helical couplings. We add the Landau damping and resonant destabilization effects using a closure relation. The simulations with multiple NBI lines show three different regimes: the non damped regime where the multi beam AEs growth rate is larger compared to the growth rate of the AEs destabilized by the individual NBI lines, the interaction regime where the multi beam AEs growth rate is smaller than the single NBI AEs and the damped regime where the AEs are suppressed. Operations in the damped regime requires EP species with different density profile flatness or gradient locations. In addition, the AEs growth rate in the interaction regime is further reduced if the combined NBI lines have similar beam temperatures and the $\beta$ of the NBI line with flatter EP density profile increases. Then, optimization trends are identified in DIII-D high poloidal $\beta$ and LHD low density / magnetic field discharges with multiple NBI lines as well as the configuration requirements to operate in the damped and interaction regimes. DIII-D simulations show a decrease of the $n=2$ to $6$ AEs growth rate and $n=1$ AE are stabilized in the LHD case. The helical coupling effects in LHD simulations lead to a transition from the interaction to the damped regime of the $n=2,-8,12$ helical family.

\end{abstract}

%Use showkeys class option if keyword

\maketitle

\section{Introduction \label{sec:introduction}}

Energetic particle (EP) driven instabilities can enhance the transport of fusion produced alpha particles, energetic neutral beams and particles heated using ion cyclotron resonance heating (ICRF) \cite{1,2,3}. The consequence is decreased heating efficiency in devices  as TFTR, JET and DIII-D tokamaks or LHD, TJ-II and W7-AS stellarators \cite{4,5,6,7,8,9}. If the mode frequency resonates with the drift, bounce or transit frequencies of the energetic particles and energy transfer occurs from particles to wave through $J \cdot E$ effects, the particle and diffusive losses increase. In addition, plasma instabilities such as internal kinks \cite{10,11,78,79}, ballooning modes \cite{12,80} and resistive wall modes \cite{81} can be kinetically destabilized/stabilized by the energetic particles.

Alfv\' en Eigenmodes (AE) are driven in the spectral gaps of the shear Alfv\' en continua \cite{13,14}, destabilized by super-Alfv\' enic alpha particles and energetic particles. Alfv\' en Eigenmode (AE) activity was observed before in several discharges and configurations \cite{15,16,17,18}. The different Alfv\' en eigenmode families ($n$ is the toroidal mode and $m$ the poloidal mode) are linked to frequency gaps produced by periodic variations of the Alfv\' en speed, for example: toroidicity induced Alfv\' en Eigenmodes (TAE) couple $m$ with $m+1$ modes \cite{19,20,21}, beta induced Alfv\' en Eigenmodes driven by compressibility effects (BAE) \cite{22}, Reversed-shear Alfv\' en Eigenmodes (RSAE) due to local maxima/minima in the safety factor $q$ profile \cite{23}, Global Alfv\' en Eigenmodes (GAE) observed in the minimum of the Alfv\' en continua \cite{24,25}, ellipticity induced Alfv\' en Eigenmodes (EAE) coupling $m$ with $m+2$ modes \cite{26,27} and noncircularity induced Alfv\' en Eigenmodes (NAE) coupling $m$ with $m+3$ or higher \cite{28,29}.

The destabilizing effect of combined EP species populations has not been extensively studied. In future nuclear fusion devices such as ITER, different EP species will coexist in the plasma, in particular NBI ions and alpha particles \cite{30,35}, so it is desirable to analyze the AE stability in these conditions. Experiments in the TFTR device already indicated that multiple EP species effects can have an important influence on AEs stability; alpha particle driven AEs were stabilized by the presence of NBI driven EP species, only measured at the end of the discharge after the beam injection was stopped \cite{36,38}. In present fusion devices the effect of AEs destabilized by alphas is absent although the combination of different NBI EP species populations could lead to similar damping effects.

High poloidal $\beta$ discharges are considered in the present study because it is a possible operational scenario for tokamak steady state operation \cite{39,40,41,42,43,44}, based on bootstrap current consistent profiles and non inductive current drive \cite{45,46,47,48,49} allowing smaller toroidal plasma currents (reduced possibility of triggering plasma disruptions), improved MHD stability (second stability regime), favorable transport properties and higher confinement factor. High poloidal $\beta$ discharges are proposed as an ITER scenario, showing a reasonable extrapolation to a reactor device size and fusion output power. We consider high poloidal $\beta$ discharges in the DIII-D plasma, heated by eight neutral beam injectors (NBI), six sources injected in the midplane (on axis) and 2 injected downwards at an angle (off axis). Six sources are injected in the direction of the plasma current (co-injected), including two tilted sources, and 2 source are injected opposite to the plasma current (counter-injected). The plasma is deuterium and the NBI also injects deuterium with a beam energy of $80$ keV ($2.25$ MW source). The destabilization of AE linked to strong NBI heating was measured before in DIII-D, triggering a large variety of AE instabilities as GAE \cite{50}, TAE \cite{51}, RSAE \cite{52}, BAE \cite{53}, EAE \cite{54} and NAE \cite{55}. The AE instabilities reduce the device performance, increasing the transport and enhancing energetic particle losses \cite{56,57,58}.

The study also includes experiments in the LHD stellarator dedicated to analyze the destabilization of AE by NBI energetic particles, easily excited in configurations with low magnetic field ($B_{0} = 0.5$ T) and bulk density ($n_{0} = 5.8 \cdot 10^{18}$ m$^{-3}$) \cite{59,60}. In this LHD configuration the plasma is heated by neutral beams injecting energetic hydrogen neutrals tangentially using three NBI lines up to 180 keV, destabilizing $n=1$ and $2$ TAE \cite{61}.

The aim of the present study is to analyze the AE stability of DIII-D high poloidal $\beta$ and LHD low density / magnetic field discharges heated by two NBI lines. If the NBI configuration leads to a decrease of the AEs growth rate compared to the AEs destabilized by the individual NBI driven EP species, we identify such NBIs operational regimes as the interaction regime. On the other hand, if the NBI configuration leads to the stabilization of the AEs although the AEs are unstable for the individual NBI driven EP species, we identify such NBIs operational regimes as the damped regime. The study consists of an NBI with a fixed configuration (identified as NBI A) along with a second NBI configuration that can be modified (identified as variable NBI B). The effects of the EP density profile, beam energy and NBI injection intensity are included in the analysis, identifying the role of the resonance of the variable NBI driven energetic particle with the thermal plasma on the growth rate and frequency of the AE destabilized by the fixed NBI driven energetic particles.

This analysis is performed using the FAR3D code \cite{62,63,64}, with extensions to include the moment equations of the energetic ion density and parallel velocity \cite{65,66} allowing treatment of linear wave-fast ion resonances. The numerical model solves the reduced non-linear resistive MHD equations adding the Landau damping/growth (wave-particle resonance effects), geodesic acoustic waves (parallel momentum response of the thermal plasma) \cite{23} and two fluid effects \cite{67}. The model requires Landau closure relations that can be calibrated by more complete kinetic models \cite{23}. The simulations are based on an equilibria calculated by the VMEC code \cite{68}.

This paper is organized as follows. The model equations, numerical scheme and equilibrium properties are described in section \ref{sec:model}. High poloidal $\beta$ discharges in DIII-D plasma with multiple NBI injection are studied in section \ref{sec:DIII-D}. Low magnetic field and bulk plasma density discharges in LHD with multiple NBI injection are analyzed in section \ref{sec:LHD}. Finally, the conclusions of this paper are presented in section \ref{sec:conclusions}.

\section{Numerical model \label{sec:model}}

A reduced set of equations to describe the evolution of the background plasma and fields, retaining the toroidal angle variation are used in the present study. These are derived from the method employed in Ref.\cite{69} assuming high-aspect ratio configurations with moderate $\beta$-values. We obtain a reduced set of equations using the exact two (tokamak) or three-dimensional (stellarator) equilibrium. The effect of the energetic particle population is included in the formulation as moments of the kinetic equation truncated with a closure relation \cite{70}. These describe the evolution of the energetic particle density ($n_{f}$) and velocity moments parallel to the magnetic field lines ($v_{||f}$). The coefficients of the closure relation are selected to match a two-pole approximation of the plasma dispersion function.    

The plasma velocity and perturbation of the magnetic field are defined as
\begin{equation}
 \mathbf{v} = \sqrt{g} R_0 \nabla \zeta \times \nabla \Phi, \quad\quad\quad  \mathbf{B} = R_0 \nabla \zeta \times \nabla \psi,
\end{equation}
where $\zeta$ is the toroidal angle, $\Phi$ is a stream function proportional to the electrostatic potential, and $\psi$ is the perturbation of the poloidal flux.

The equations, in dimensionless form, are
\begin{equation}
\label{poloidal flux}
\frac{\partial \tilde \psi}{\partial t} =  \sqrt{g} B \nabla_{||} \Phi  + \eta \varepsilon^2 J \tilde J^\zeta + \rlap{-}\iota \frac{\Delta S \beta_{0}}{2\varepsilon^2 \omega_{cy} \sqrt{g}} \nabla_{||} p
\end{equation}
\begin{eqnarray} 
\label{vorticity}
\frac{{\partial \tilde U}}{{\partial t}} =  -\epsilon v_{\zeta,eq} \frac{\partial U}{\partial \zeta} \nonumber\\
+ S^2 \left[{ \sqrt{g} B \nabla_{||} J^\zeta - \frac{\beta_0}{2\varepsilon^2} \sqrt{g} \left( \nabla \sqrt{g} \times \nabla \tilde p \right)^\zeta }\right]   \nonumber\\
-  S^2 \left[{\frac{\beta_f^{A,B}}{2\varepsilon^2} \sqrt{g} \left( \nabla \sqrt{g} \times \nabla \tilde n_f^{A,B} \right)^\zeta }\right] \nonumber\\
+ \frac{S \beta_{0} (1 - \Delta)}{2 \omega_{cy} \varepsilon^2} \left[\nabla \times \left( \sqrt{g} \left((B \times \nabla p) \cdot \nabla \right) v_{\perp}     \right) \right]^{\zeta}
\end{eqnarray} 
\begin{eqnarray}
\label{pressure}
\frac{\partial \tilde p}{\partial t} = -\epsilon v_{\zeta,eq} \frac{\partial p}{\partial \zeta} + \frac{dp_{eq}}{d\rho}\frac{1}{\rho}\frac{\partial \tilde \Phi}{\partial \theta} \nonumber\\
 +  \Gamma p_{eq}  \left[{ \sqrt{g} \left( \nabla \sqrt{g} \times \nabla \tilde \Phi \right)^\zeta - \nabla_{||}  v_{|| th} }\right] \nonumber\\
 + \frac{S \beta_{0} (1 - \Delta)}{2\varepsilon^2 \omega_{cy} \sqrt{g}} p \nabla p \cdot (\nabla \times B)
\end{eqnarray} 
\begin{eqnarray}
\label{velthermal}
\frac{{\partial \tilde v_{|| th}}}{{\partial t}} = -\epsilon v_{\zeta,eq} \frac{\partial v_{||th}}{\partial \zeta} -  \frac{S^2 \beta_0}{n_{0,th}} \nabla_{||} p 
\end{eqnarray}
\begin{eqnarray}
\label{nfast}
\frac{{\partial \tilde n_f}}{{\partial t}}^{A,B} = -\epsilon v_{\zeta,eq} \frac{\partial n_{f}^{A,B}}{\partial \zeta} - \frac{S  (v_{th,f}^{A,B})^2}{\omega_{cy}^{A,B}} \Omega_d (\tilde n_f^{A,B})  \nonumber\\
- S  n_{f0}^{A,B} \nabla_{||} v_{\| f}^{A,B} - \varepsilon^2  n_{f0}^{A,B} \Omega_d (\tilde \Phi) + \varepsilon^2 n_{f0}^{A,B} \Omega_{*} (\tilde  \Phi) 
\end{eqnarray}
\begin{eqnarray}
\label{vfast}
\frac{{\partial \tilde v_{|| f}}}{{\partial t}}^{A,B} = -\epsilon v_{\zeta,eq} \frac{\partial v_{||f}^{A,B}}{\partial \zeta}  -  \frac{S  (v_{th,f}^{A,B})^2}{\omega_{cy}^{A,B}} \, \Omega_d (\tilde v_{|| f}^{A,B}) \nonumber\\
- \left( \frac{\pi}{2} \right)^{1/2} S  v_{th,f}^{A,B} \left| \nabla_{||} \right|  v_{|| f}^{A,B}   \nonumber\\
- \frac{S  (v_{th,f}^{A,B})^2}{n_{f0}^{A,B}} \nabla_\| n_{f}^{A,B} + S \varepsilon^2  (v_{th,f}^{A,B})^2 \Omega_* (\tilde \psi) 
\end{eqnarray}
The components from EP particles species A and B are coupled with the thermal plasma through the third terms of the vorticity equation (Eq. 3). Here, $U =  \sqrt g \left[{ \nabla  \times \left( {\rho _m \sqrt g {\bf{v}}} \right) }\right]^\zeta$ is the vorticity and $\rho_m$ the ion and electron mass density. The perturbed toroidal current density $J^{\zeta}$ is defined as:
\begin{eqnarray}
J^{\zeta} =  \frac{1}{\rho}\frac{\partial}{\partial \rho} \left(-\frac{g_{\rho\theta}}{\sqrt{g}}\frac{\partial \psi}{\partial \theta} + \rho \frac{g_{\theta\theta}}{\sqrt{g}}\frac{\partial \psi}{\partial \rho} \right) \nonumber\\
- \frac{1}{\rho} \frac{\partial}{\partial \theta} \left( \frac{g_{\rho\rho}}{\sqrt{g}}\frac{1}{\rho}\frac{\partial \psi}{\partial \theta} + \rho \frac{g_{\rho \theta}}{\sqrt{g}}\frac{\partial \psi}{\partial \rho} \right)
\end{eqnarray}
$v_{||th}$ is the parallel velocity of the thermal particles, $v_{\zeta,eq}$ is the equilibrium toroidal rotation and $v_{\perp} = - \nabla \Phi \times B^{\zeta} e_{\zeta} / B^2$ is the thermal perpendicular velocity. $n_{f}$ is normalized to the density at the magnetic axis $n_{f_{0}}$, $\Phi$ to $a^2B_{0}/\tau_{R}$ and $\Psi$ to $a^2B_{0}$. All lengths are normalized to a generalized minor radius $a$; the resistivity to $\eta_0$ (its value at the magnetic axis); the time to the resistive time $\tau_R = a^2 \mu_0 / \eta_0$; the magnetic field to $B_0$ (the averaged value at the magnetic axis); and the pressure to its equilibrium value at the magnetic axis. The Lundquist number $S$ is the ratio of the resistive time to the Alfv\' en time $\tau_{A0} = R_0 (\mu_0 \rho_m)^{1/2} / B_0$. $\rlap{-} \iota$ is the rotational transform, $v_{th,f} = \sqrt{T_{f}/m_{f}}/v_{A0}$ the energetic particle thermal velocity normalized to the Alfv\' en velocity in the magnetic axis and $\omega_{cy}$ the energetic particle cyclotron frequency times $\tau_{A0}$. $q_{f}$ is the charge, $T_{f}$ the temperature and $m_{f}$ the mass of the energetic particles. $\Delta$ is the electron pressure normalized to the total pressure. The $\Omega$ operators are defined as:
\begin{eqnarray}
\label{eq:omedrift}
\Omega_d = \frac{1}{2 B^4 \sqrt{g}}  \left[  \left( \frac{I}{\rho} \frac{\partial B^2}{\partial \zeta} - J \frac{1}{\rho} \frac{\partial B^2}{\partial \theta} \right) \frac{\partial}{\partial \rho}\right] \nonumber\\
-   \frac{1}{2 B^4 \sqrt{g}} \left[ \left( \rho \beta_* \frac{\partial B^2}{\partial \zeta} - J \frac{\partial B^2}{\partial \rho} \right) \frac{1}{\rho} \frac{\partial}{\partial \theta} \right] \nonumber\\ 
+ \frac{1}{2 B^4 \sqrt{g}} \left[ \left( \rho \beta_* \frac{1}{\rho} \frac{\partial B^2}{\partial \theta} -  \frac{I}{\rho} \frac{\partial B^2}{\partial \rho} \right) \frac{\partial}{\partial \zeta} \right]
\end{eqnarray}

\begin{eqnarray}
\label{eq:omestar}
\Omega_* = \frac{1}{B^2 \sqrt{g}} \frac{1}{n_{f0}} \frac{d n_{f0}}{d \rho} \left( \frac{I}{\rho} \frac{\partial}{\partial \zeta} - J \frac{1}{\rho} \frac{\partial}{\partial \theta} \right) 
\end{eqnarray}
Here the $\Omega_{d}$ operator is constructed to model the average drift velocity of a passing particle and $\Omega_{*}$ models its diamagnetic drift frequency. We also define the parallel gradient and curvature operators:
\begin{equation}
\label{eq:gradpar}
\nabla_\| f = \frac{1}{B \sqrt{g}} \left( \frac{\partial \tilde f}{\partial \zeta} +  \rlap{-} \iota \frac{\partial \tilde f}{\partial \theta} - \frac{\partial f_{eq}}{\partial \rho}  \frac{1}{\rho} \frac{\partial \tilde \psi}{\partial \theta} + \frac{1}{\rho} \frac{\partial f_{eq}}{\partial \theta} \frac{\partial \tilde \psi}{\partial \rho} \right)
\end{equation}
\begin{equation}
\label{eq:curv}
\sqrt{g} \left( \nabla \sqrt{g} \times \nabla \tilde f \right)^\zeta = \frac{\partial \sqrt{g} }{\partial \rho}  \frac{1}{\rho} \frac{\partial \tilde f}{\partial \theta} - \frac{1}{\rho} \frac{\partial \sqrt{g} }{\partial \theta} \frac{\partial \tilde f}{\partial \rho}
\end{equation}
with the Jacobian of the transformation:
\begin{equation}
\label{eq:Jac}
\frac{1}{\sqrt{g}} = \frac{B^2}{\varepsilon^2 (J+ \rlap{-} \iota I)}
\end{equation}

Equations~\ref{pressure} and~\ref{velthermal} introduce the parallel momentum response of the thermal plasma, required for coupling to the geodesic acoustic waves, accounting the geodesic compressibility in the frequency range of the geodesic acoustic mode (GAM) \cite{71,72}. The last term in the equations~\ref{poloidal flux},~\ref{vorticity} and~\ref{pressure} introduce the two fluid effects adding the diamagnetic currents in the thermal plasma components \cite{67}. The index A and B indicate that the model includes equations for the EP density and parallel velocity of two separate NBI driven EP species. The EP species of the model are treated as independent uncoupled populations with separate density and parallel velocity momentum equations, interacting only through the fields $\tilde \Phi$ and $\tilde \psi $. 

Equilibrium flux coordinates $(\rho, \theta, \zeta)$ are used. Here, $\rho$ is a generalized radial coordinate proportional to the square root of the toroidal flux function, and normalized to one at the edge. The flux coordinates used in the code are those described by Boozer \cite{61}, and $\sqrt g$ is the Jacobian of the coordinate transformation. All functions have equilibrium and perturbation components represented as: $ A = A_{eq} + \tilde{A} $. 

The FAR3D code uses finite differences in the radial direction and Fourier expansions in the two angular variables. The numerical scheme is semi-implicit in the linear terms. The nonlinear version uses a two semi-step method to ensure $(\Delta t)^2$ accuracy.

The present model was already used to study the AE activity in LHD \cite{60,61}, TJ-II \cite{73,74,75} and DIII-D \cite{76} indicating reasonable agreement with the observations.

\subsection{Equilibrium properties}

We use fixed boundary results from the VMEC equilibrium code \cite{68} calculated using the DIII-D reconstruction of the high poloidal $\beta$ discharge 166495 at $t=3650$ ms and low density /magnetic field LHD discharge 41503. 

\begin{figure}[h]
\centering
\includegraphics[width=1\columnwidth]{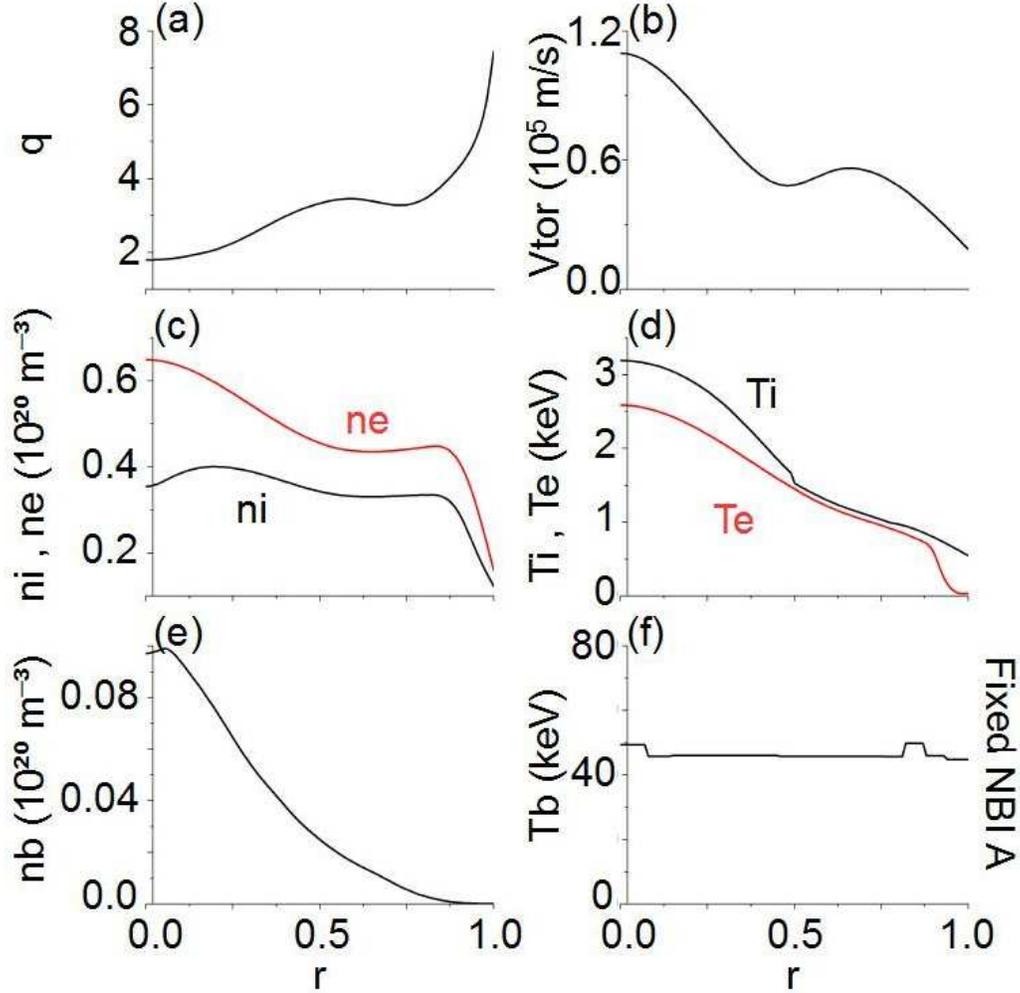} 
\caption{DIII-D profiles: (a) q profile, (b) toroidal rotation, (c) electron and ion density, (d) electron and ion temperature, fixed NBI A EP density (e) and temperature (f).}
\label{FIG:1}
\end{figure}

The experimental constraints used in the DIII-D equilibrium reconstruction are taken from magnetic data, MSE data, kinetic pressure and edge density profile from NEO model. Due to the fact that the FAR3D stability model is based on stellarator symmetry, we null out the up-down asymmetric terms in the VMEC shape and base the calculations for the current paper on up-down symmetric equilibria. Since the original DIII-D experiments were run in single-null divertor mode, the equilibria we use here will be nearby, but slightly different from the experimental ones. The consequence is a little displacement of the flux/magnetic surfaces and a small variation of the modes growth rate and frequency, although the plasma stability properties are almost the same. The magnetic field at the magnetic axis is $2$ T, the averaged inverse aspect ratio is $\varepsilon=0.47$ and $\beta_{0}$ is $5.7 \%$ \cite{77}. The energy of the injected particles by the fixed NBI (NBI A) is $T_{b}(0) = 49.32$ keV ($v_{th,f} = 2.173 \cdot 10^{6}$ m/s). Figure~\ref{FIG:1} shows the thermal plasma and fixed NBI EP profiles in the DIII-D discharge.

\begin{figure}[h]
\centering
\includegraphics[width=1\columnwidth]{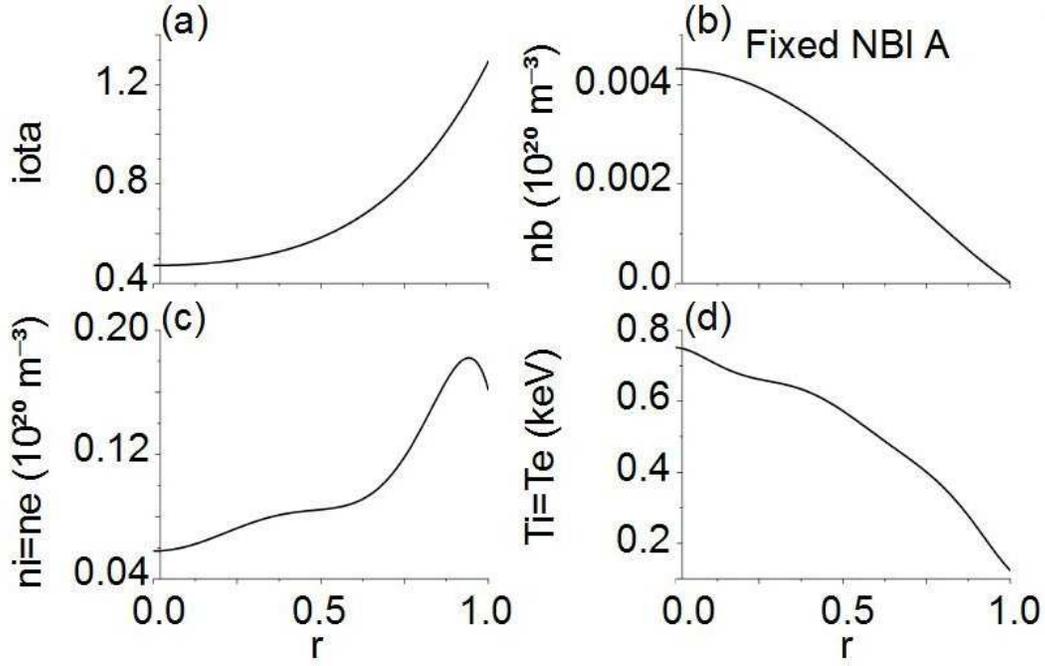} 
\caption{LHD profiles: (a) iota profile, (b) fixed NBI A EP density, (c) electron and ion density, (d) electron and ion temperature.}
\label{FIG:2}
\end{figure}

In the LHD equilibrium, the electron density and temperature profiles were reconstructed by Thomson scattering data and electron cyclotron emission. The vacuum magnetic axis is inward-shifted with $R_{\rm{axis}} = 3.76$ m. The magnetic field at the magnetic axis is $0.619$ T, the inverse aspect ratio $\varepsilon$ is $0.15$ and $\beta_0$ is $4.2 \%$. The injection energy of the fixed NBI is $T_{b} = 180$ KeV but we nominally consider only $100$ keV (energetic particle thermal velocity of $v_{th,f} = 3.1 \cdot 10^6$ m/s), resulting in an averaged Maxwellian energy equal to the average energy of a slowing-down distribution with 180 keV. In this case the EP energy is considered constant, with no radial variation, due to the lack of $T_{b}$ experimental or modeling data. Figure~\ref{FIG:2} shows the thermal plasma and fixed NBI EP profiles for the LHD discharge.

\begin{figure}[h!]
\centering
\includegraphics[width=0.5\textwidth]{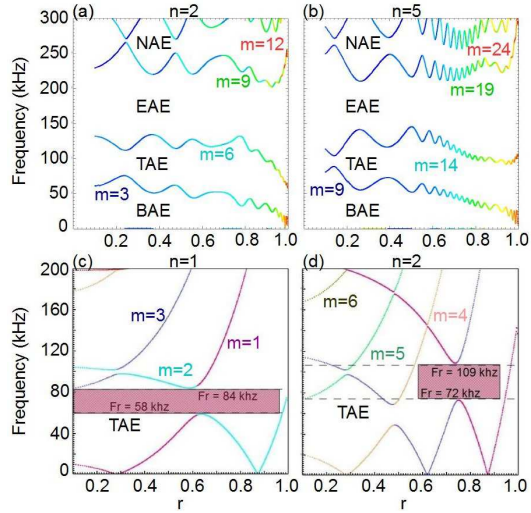}
\caption{Alfv\' en gaps in DIII-D shot 166495 at $t=3650$ ms for $n=2$ (a) and $n=5$ (b). Alfv\' en gaps in LHD shot 41503 for $n=1$ (c) and $n=2$ (d). The analysis only takes account of the lowest toroidal mode families $n=1$ and $2$.}\label{FIG:3}
\end{figure}

Figure~\ref{FIG:3} shows the Alfv\' en gaps in the DIII-D discharge for $n=2$ and $5$ toroidal modes as well as in the LHD discharge for $n=1$ and $2$ toroidal modes. In the DIII-D case there are four main Alfv\' en gaps: TAE gap between $[50,120]$ kHz, EAE gap between $[120,210]$ kHz and NAE gap for $f>210$ Khz. BAE, BAAE and GAE are destabilized below $f= 50$ kHz. In LHD case $n=1$ TAEs are destabilized between $[58,84]$ kHz and $n=2$ TAEs between $[72,109]$ kHz.

\subsection{Simulation parameters}

The simulations are performed with a uniform radial grid of 1000 points. The dynamic and equilibrium toroidal (n) and poloidal (m) modes included in the study are summarized in table~\ref{Table:1} for DIII-D and LHD cases. The toroidal modes $n=8$ to $12$ are included only in the LHD simulations with helical couplings. In the following, the mode number notation is $m/n$ in the section where the DIII-D discharge is analyzed, consistent with the $q=m/n$ definition for the associated rational surface. On the other hand, in the LHD section the mode number notation is $n/m$ consistent with an $\rlap{-}\iota=n/m$ rational surface location.

\begin{table}[h]
\centering

\begin{tabular}{c}
DIII-D \\
\end{tabular}

\begin{tabular}{c | c }
\hline
Dy (n) & Dy (m) \\ \hline
$1$ & $[2,5]$ \\
$2$ & $[4,10]$  \\
$3$ & $[6,15]$  \\
$4$ & $[8,18]$  \\
$5$ & $[10,20]$  \\
$6$ & $[12,24]$  \\ \hline
Eq (n) & Eq (m)  \\ \hline
$0$ & $[0,9]$ \\ \hline
\end{tabular}

\hspace{2cm}

\begin{tabular}{c}
LHD \\
\end{tabular}

\begin{tabular}{c | c }
\hline
Dy (n) & Dy (m)  \\ \hline
$1$ & $[1,8]$  \\
$2$ & $[2,12]$  \\
$8$ & $[5,15]$  \\
$9$ & $[6,18]$  \\
$11$ & $[6,22]$  \\
$12$ & $[7,24]$  \\
Eq (n) & Eq (m) \\ \hline
$0$ & $[0,4]$ \\
$10$ & $[-7,3]$ \\ \hline
\end{tabular}

\caption{Dynamic and equilibrium toroidal (n) and poloidal (m) modes in the simulation of DIII-D and LHD cases.} \label{Table:1}
\end{table}

The kinetic closure moment equations (6) and (7) break the usual MHD parities. This is taken into account by including both parities $sin(m\theta + n\zeta)$ and $cos(m\theta + n\zeta)$ for all dynamic variables, and allowing for both a growth rate and real frequency in the eigenmode time series analysis. The convention of the code is, in case of the pressure eigenfunction, that $n > 0$ corresponds to the Fourier component $\cos(m\theta + n\zeta)$ and $n < 0$ to $\sin(-m\theta - n\zeta)$. For example, the Fourier component for mode $-7/2$ is $\cos(-7\theta + 2\zeta)$ and for the mode $7/-2$ is $\sin(-7\theta + 2\zeta)$. The magnetic Lundquist number is $S=5\cdot 10^6$ similar to the experimental value in the middle of the plasma.

The density ratio between the energetic particles and bulk plasma ($n_{f}(0)/n_{e}(0)$) at the magnetic axis is controlled through the $\beta_{f}=$ value, linked to the NBI injection intensity, calculated for the DIII-D case by the code TRANSP without the effect of the anomalous beam ion transport. The ratio between the energetic particle thermal velocity and Alfv\' en velocity at the magnetic axis ($v_{th,f}/v_{A0}$) controls the resonance coupling efficiency between AE and energetic particles, associated with the NBI voltage or beam energy. We consider a Maxwellian distribution for the energetic particle distribution function.

\section{Multiple NBI lines in DIII-D high poloidal $\beta$ discharges \label{sec:DIII-D}}

\begin{figure}[h!]
\centering
\includegraphics[width=0.45\textwidth]{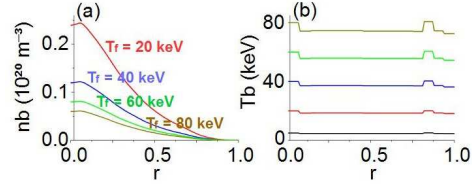}
\caption{Density (a) and temperature (b) profiles of the variable NBI B for the DIII-D case simulations.}
\label{FIG:4}
\end{figure}

\begin{figure}[h!]
\centering
\includegraphics[width=0.45\textwidth]{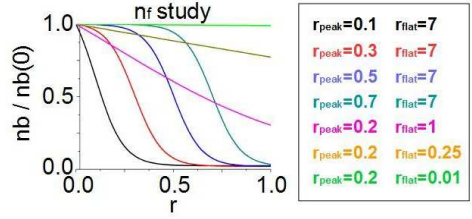}
\caption{Density profiles of the variable NBI B in the study where the EP density distribution is modified. These are used in both DIII-D and LHD case simulations. The variable $r_{flat}$ controls the the profile gradient and $r_{peak}$ the location of the gradient maximum along the normalized minor radius.}
\label{FIG:5}
\end{figure}

In this section we study the effect of multiple NBI lines in DIII-D high poloidal $\beta$ discharges, identifying the optimal configuration of the variable NBI to minimize the AE growth rate. First we study the effect of the variable NBI beam temperature and $\beta_{f}$, then the effect of the EP density profile. The profiles of the variable NBI profiles are shown in figures~\ref{FIG:4} and \ref{FIG:5}. The analytic expression used for the EP density profile is the following:

$$n_{b}(r) = \frac{(0.5 (1+ \tanh(r_{flat} \cdot (r_{peak}-r))+0.02)}{(0.5 (1+\tanh(r_{flat} \cdot r_{peak}))+0.02)}$$
The location of the gradient is controlled by the parameter ($r_{peak}$) and the flatness by ($r_{flat}$).

The study of the variable NBI beam energy is performed keeping $\beta_{f} = 0.0464$, the same $\beta_{f}$ as the fixed NBI component. In the study of the variable NBI $\beta_{f}$, $T_{b} = 40$ and $80$ keV are used. The variable NBI EP density in each simulation is consistent with the expression: $n_{b} = \beta_{f} B_{0}^{2} / 2 T_{b} \mu_{0} k_{B}$. $k_{B}$ is the Boltzmann constant.

\subsection{Effect of the variable NBI beam temperature and injection intensity}

\begin{figure}[h!]
\centering
\includegraphics[width=0.5\textwidth]{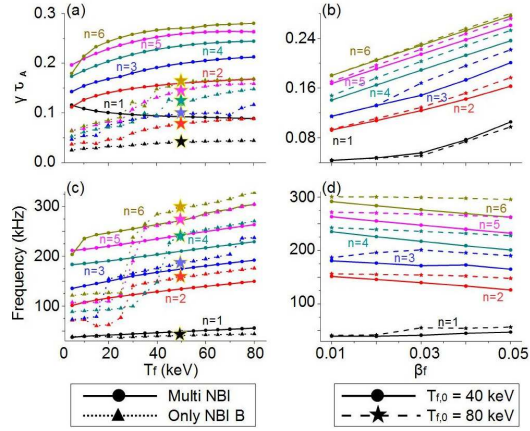}
\caption{AE (a) growth rate and (c) frequency in the study where the variable NBI B beam temperature is modified. The solid lines show the multiple NBI simulations, the dotted lines the single variable NBI B simulations and the stars the single fixed NBI A simulations. AE (b) growth rate and (d) frequency in the study where the variable NBI B $\beta_{f}$ is modified. The solid lines show the multiple NBI simulations with $T_{b} = 40$ keV and the dashed lines with $T_{b} = 80$ keV.}
\label{FIG:6}
\end{figure}

Figure~\ref{FIG:6} shows the AE growth rate ($\gamma$) and frequency (f) if the variable NBI B beam temperature and $\beta_{f}$ are modified. The solid lines show the simulations with multiple NBI, the dotted lines the simulations with only the variable NBI and the stars the simulations with only the fixed NBI. The AE growth rate and frequency change if the variable NBI beam temperature is modified (see panel a and c). In particular, the growth rate of the $n=2-6$ AEs increases if the variable NBI temperature increases, decreasing for the $n=1$ AE. If we compare the growth rate of the multiple beam simulations with the simulations with only the fixed or the variable NBI, the AE destabilized by the combined beams shows a larger growth rate, so no damping effect exist. On the other hand, the different tendencies of the $n=1$ profile in the multiple beam simulations (negative slope) regarding the simulations with only the variable NBI (positive slope), suggesting that the resonance characteristics of the variable NBI affects the properties of the AE, leading to a lower growth rate. Consequently, it should be possible to find a configuration where the resonance properties of the variable NBI leads to AE with lower growth rate than the AE destabilized by a single NBI, in a manner that the variable NBI will drive a stabilizing effect over the perturbation caused by the fixed NBI. If we analyze the dependency of the AE frequency with the variable NBI beam temperature we observe an increase and increment of the AE frequency with the beam temperature. It should be noted that the profile tendency of the multiple beam regarding the single beam simulations is different for the $n=2-6$ AE, showing a sharp increase for the single NBI simulations above a specific temperature while the profile slope is almost constant for the multiple NBI simulations, pointing out a transition between different families of AEs, identified as an increase of the AE growth rate and frequency. These transitions where already observed and analyzed in previous studies, linked with an enhancement of the energetic particle forcing caused by an improved resonance efficiency between the energetic particles and bulk plasma \cite{61,75}. The transition is not observed in the multiple beam simulations because the combined effect of both NBIs is strong enough to destabilize the AE family with higher growth rate and frequency. In the second part of the study we analyze the AE growth rate if the $\beta_{f}$ of the variable NBI is modified (see panel b and d) if the NBI B beam temperature is $T_{b} = 40$ keV (solid lines) or $T_{b} = 80$ keV (dashed lines). The enhancement of the variable NBI deposition intensity leads to an increase of the AE growth rate and a drop of the frequency for all modes, pointing out that the variable NBI effect leads to an enhancement of the fixed NBI perturbation.

\begin{figure}[h!]
\centering
\includegraphics[width=0.5\textwidth]{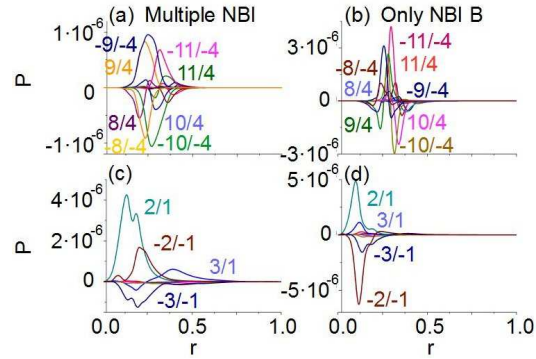}
\caption{Pressure eigenfunction of $n=4$ AE if $T_{b} = 20$ keV for multiple (a) and single (b) NBI simulations. Pressure eigenfunction of $n=1$ AE if $T_{b} = 80$ keV for multiple (c) and single (d) NBI simulations.}
\label{FIG:7}
\end{figure}

Figure~\ref{FIG:7} shows the pressure eigenfunctions of the $n=4$ AE if $T_{b} = 20$ keV for multiple (a) and single (b) NBI simulations as well as the  $n=1$ AE if $T_{b} = 80$ keV for multiple (c) and single (d) NBI simulations.  The $n=4$ AEs in the single NBI simulations is a TAE destabilized in the inner plasma region by the coupled $11/4$ and $10/4$ modes. In the multiple NBI simulations the eigenfunction width is wider and the toroidal coupling is enhanced, leading to the destabilization of a EAE/NAE by the modes $8/4$ to $11/4$. The transition from the TAE to the EAE/NAE is caused by the enhancement of the fixed NBI perturbation by the variable NBI. The $n=1$ AEs are BAEs destabilized nearby the magnetic axis by $2/1$ mode. Again the eigenfunction width is larger in the multiple NBI simulation due to the enhancement of the perturbation, although the destabilization of the AE is weaker compared to the $n=4$ so no transition between AE families is observed, because the resonance of the variable NBI with the bulk plasma if $T_{b} = 80$ keV is less efficient, pointing out the essential role of the EP resonance in the multiple beam simulations.

In summary, if the temperature of the variable NBI is modified the EP resonance with the bulk plasma changes, as well as the growth rate and frequency of the AEs in the multiple beam simulations. Consequently, there are configurations of the variable NBI that lead to a weaker destabilization of the AE driven by the fixed NBI. The next step of the study consists in analyzing the effect of the EP density profile of the variable NBI on the AE growth rate, with the aim to identify NBI operational regimes with multiple beam damping effects. 

\subsection{Effect of the variable NBI driven EP density profile}

\begin{figure}[h!]
\centering
\includegraphics[width=0.5\textwidth]{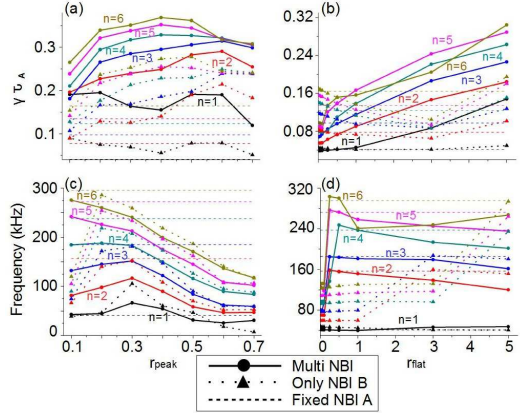}
\caption{AE growth rate (a) and frequency (c) in the study where the deposition region of the variable NBI is modified ($r_{peak}$). AE growth rate (b) and frequency (d) in the study where the flatness of the variable NBI driven EP density profile is modified ($r_{flat}$). The solid lines show the multiple NBI simulations, the dotted lines the single variable NBI B simulations and the dashed lines the single fixed NBI A simulations.}
\label{FIG:8}
\end{figure}

Figure~\ref{FIG:8} shows the AE growth rate and frequency for different configurations of the EP density profile of the variable NBI, where NBI B $T_{b} = 40$ keV and $\beta_{f} = 0.0464$. The solid lines show the multiple NBI simulations, the dotted lines the single variable NBI simulations and the dashed line the single fixed NBI simulations. If the location of the EP density profile gradient is modified (panels a and c), the multiple NBI simulations show a larger growth rate compared to single NBI simulations for all the deposition regions analyzed, so no multiple beam damping effects are observed. The profiles in the multiple and single NBI simulations show similar trends, the growth rate decreases if the NBI is deposited on-axis, showing a local maximum for off-axis NBI depositions in the middle of the plasma.

The study of the EP density profile flatness (panels b and d) indicates that the AE growth rate in the multiple NBI simulations is smaller compared to the single NBI simulations if $r_{flat} < 0.5$, so there is an stabilizing effect of the variable NBI over the AEs destabilized by the fixed NBI. In addition, the AE frequency in the multiple NBI simulations is smaller compared to the simulation with only the fixed NBI and similar to simulations with the variable NBI if $r_{flat} < 0.5$. In the following, we define the multiple NBI configurations with damping effects (multiple AEs growth rate smaller than the single AEs growth rate) as ''the interaction regime''.

\begin{figure}[h!]
\centering
\includegraphics[width=0.5\textwidth]{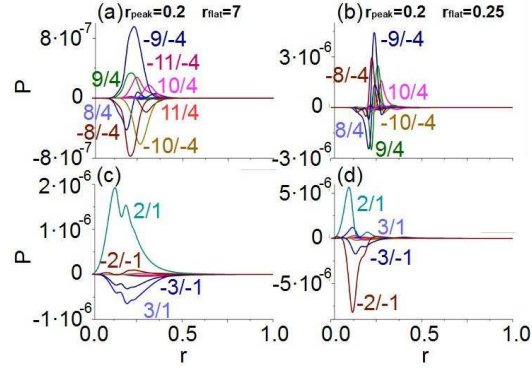}
\caption{Pressure eigenfunctions of $n=1$ for  $r_{peak} = 0.2$ if (a) $r_{flat}=7$ and if (b) $r_{flat}=0.25$. Pressure eigenfunctions of $n=4$ for  $r_{peak} = 0.2$ if (c) $r_{flat}=7$ and if (d) $r_{flat}=0.25$.}
\label{FIG:9}
\end{figure}

Figure~\ref{FIG:9} shows the pressure eigenfunctions of $n=1$ and $4$ in the non damped (panels a and c) and interaction regimes (panels b an d). The eigenfunctions width is smaller in the interaction regime. In addition, the $n=4$ shows a transition from a $8/4-11/4$ EAE/NAE in the non damped regime to a $8/4-9/4$ TAE in the interaction regime, pointing out a weaker toroidal mode coupling. Both are the consequences of the weaker EP driving in the interaction regime.

Having identified the variable NBI configuration that leads to a stabilizing effect over the fixed NBI perturbation, we analyze again the effect of the variable NBI beam temperature and deposition intensity on the AE growth rate and frequency, although this time for the interaction regime.

\subsection{Effect of the variable NBI beam temperature and injection intensity in the interaction regime}

\begin{figure}[h!]
\centering
\includegraphics[width=0.5\textwidth]{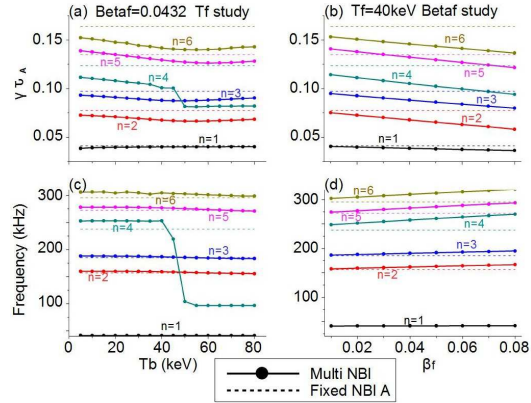}
\caption{AE growth rate (a) and frequency (c) if the variable NBI B beam temperature is modified in the interaction regime. AE (b) growth rate and (d) frequency if the variable NBI B $\beta_{f}$ is modified in the interaction regime. The solid lines show the multiple NBI simulations and the dashed lines the single fixed NBI A simulations.}
\label{FIG:10}
\end{figure}

Figure~\ref{FIG:10} shows the AE growth rates and frequencies in the interaction regime ($r_{flat}=0.1$ and $r_{peak}=0.2$) if the variable NBI beam temperature  (fixed $\beta_{f} = 0.232$), panels a and c, or $\beta_{f}$ (fixed $T_{b} = 40$ keV), panels b and d, are modified. The damping effect of the variable NBI is stronger, leading to a local minimum of the AE growth rate, if the beam temperature is similar to the fixed NBI. The decrease of the $n=4$ AE growth rate is larger because there is a transition from an EAE/NAE with $f \approx 250$ kHz if $T_{b} \le 40$ keV to a TAE with $f \approx 100$ kHz if $T_{b} > 40$ keV  (panel c). The frequency of the rest of the AEs is similar for all $T_{b}$ values. Regarding the variable NBI injection intensity, the AE growth rate decreases as $\beta_{f}$ increases, pointing out that the stabilizing effect is reinforced if the $\beta_{f}$ increases. In addition, the AEs frequency slightly increases as $\beta_{f}$ increases.

In summary, for a DIII-D high poloidal $\beta$ discharge with multiple NBI lines operating in the non damped regime, the simulations suggest a reinforcement of the EP perturbation if the variable NBI beam temperature or the injection intensity increases (except for the $n=1$ mode). On the other hand, if the multiple NBI lines operate in the interaction regime, observed if the variable NBI density profile is flatter than the fixed NBI ($r_{flat} < 0.5$), the damping effect is enhanced if both beam line temperatures are similar and the variable NBI injection intensity increases.

\section{Multiple NBI lines in LHD low density and magnetic field discharges \label{sec:LHD}}

\begin{figure}[h!]
\centering
\includegraphics[width=0.3\textwidth]{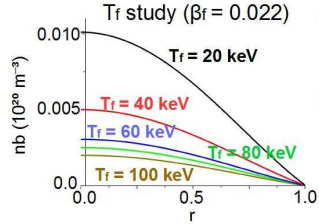}
\caption{ EP density profiles of the variable NBI for the LHD case.}
\label{FIG:11}
\end{figure}

In this section we analyze the effect of multiple NBI components in LHD low density / magnetic field discharges. We use the same framework of the previous section. The density profiles of the variable NBI EP used in the study are summarized in figure~\ref{FIG:11}. The variable NBI $T_{b}$ is constant (no radial variation).

\subsection{Effect of the variable NBI EP density profile}

\begin{figure}[h!]
\centering
\includegraphics[width=0.5\textwidth]{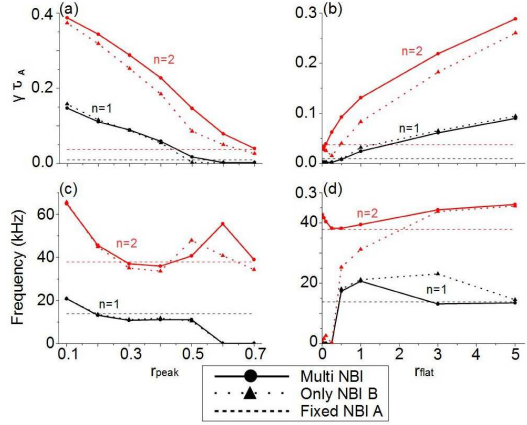}
\caption{AE growth rate (a) and frequency (c) in the study where the density profile  of the variable NBI EP is modified by changing the deposition region ($r_{peak}$). The AE growth rate (b) and frequency (d) are shown if the density profile of the variable NBI driven EP flatness is modified ($r_{flat}$). The solid lines show the multiple NBI simulations, the dotted lines the single variable NBI B simulations and the dashed lines the single fixed NBI A simulations.}
\label{FIG:12}
\end{figure}

Figure~\ref{FIG:12} shows the $n=1$ and $2$ AE growth rates and frequencies in the studies where the density profile of the variable NBI EP is modified (variable NBI $T_{b,0} = 48$ keV and $\beta_{f} = 0.0212$). The multiple NBI damping effects stabilize the $n=1$ AE if $r_{peak} > 0.5$ or $r_{flat} < 0.5$. These multiple NBI operational regimes are defined as a ''damped regime''. On the other hand, the $n=2$ AE growth rate in the multiple beam simulations is above the single NBI simulations (non damped regime). The AEs growth rate decreases in the multiple and single NBI B simulations if $r_{peak}$ increases, so an on-axis NBI deposition leads to the most unstable configuration. The growth rate of the $n=1$ ($n=2$) AE is lower in the multiple beam configuration compared to the single NBI cases if $r_{peak}$ is between $0.3$ and $0.5$ ($0.3$ and $0.4$) or $r_{flat}$ is between $0.5$ and $3$, so the multiple NBI configuration is in the interaction regime. The AEs frequency decreases if the variable NBI is deposited off-axis, except for the $n=2$ AE showing a local minimum if $r_{peak}=0.4$. In addition, the $n=1$  ($n=2$) AE frequency increases (decreases) as the density profile of the variable NBI EP is flattened, except if $r_{flat}<1$ ($r_{flat}<0.5$).

\begin{figure}[h!]
\centering
\includegraphics[width=0.5\textwidth]{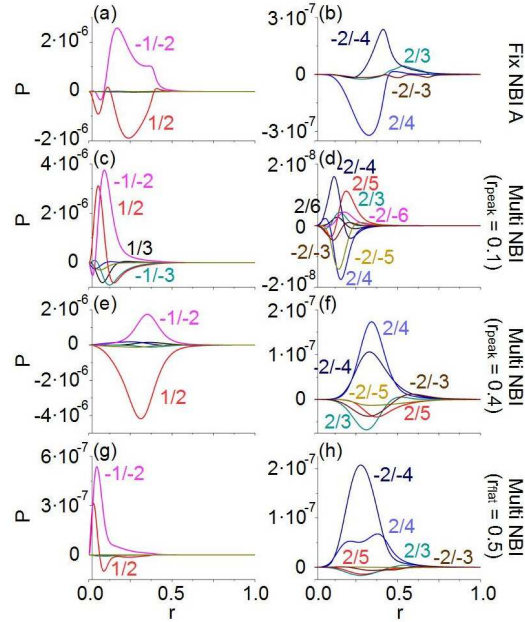}
\caption{Pressure eigenfunctions in fixed NBI A simulations for $n=1$ (a) and $n=2$ (b) AEs. Multiple NBI simulations with on-axis variable NBI B deposition ($r_{peak}=0.1$) for $n=1$ (c) and $n=2$ (d) AEs. Multiple NBI simulations with off-axis variable NBI B deposition ($r_{peak}=0.4$) for $n=1$ (e) and $n=2$ (f) AEs. Multiple NBI simulations with a flat variable NBI B EP density profile ($r_{flat}=0.5$) for $n=1$ (g) and $n=2$ (h) AEs.}
\label{FIG:13}
\end{figure}

Figure~\ref{FIG:13} shows the pressure eigenfunctions of $n=1$ and $2$ AEs in the multiple and single NBI simulations. For the fixed NBI A simulations, $1/2$ and $2/4$ BAE are destabilized in the inner plasma region (panels a and b). If a second NBI line is deposited on-axis (non damped regime), an $1/2-1/3$ TAE (panel c) and an $2/3-2/5$ EAE (panel d) are destabilized in the inner plasma. If the second NBI line is deposited off-axis (interaction regime), a $1/2$ BAE (panel e) and a $2/3-2/5$ EAE (panel f) are destabilized in the inner plasma region. If the second NBI line is deposited in the inner plasma region ($r_{peak} = 0.1$) and the EP density profile is flatter than the fixed NBI, an $n=1/2$ and a $n=2/4$ BAE are destabilized near the magnetic axis and in the inner plasma, respectively. If we compare the eigenfunction structure of the non damped (panel c) and interaction regimes (panel g), we observed a weaker EP driving in the interaction regime leading to narrower eigenfunctions.

\begin{figure}[h!]
\centering
\includegraphics[width=0.5\textwidth]{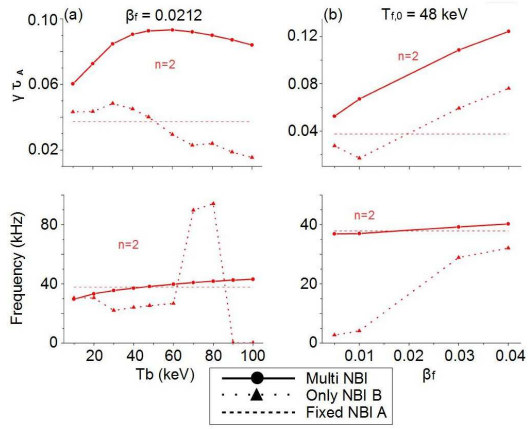}
\caption{AE growth rate (a) and frequency (c) if the variable NBI B beam temperature is modified in the non damped regime for $n=2$. AE (b) growth rate and (d) frequency if the variable NBI B $\beta_{f}$ is modified in the non damped regime for $n=2$. The solid lines show the multiple NBI simulations, the dotted lines the simulations with only the variable NBI and the dashed lines the single fixed NBI A simulations.}
\label{FIG:14}
\end{figure}

Consequently, the multiple beam damping effects are strong enough to reduce the growth rate of the $n=1$ AE, although no multiple beam damping is observed for the $n=2$ AE. If we analyze the effect of the beam temperature and injection intensity in the resonance properties of the $n=2$ AE in the configurations with the lowest growth rate for the multiple beam simulations ($r_{peak} = 0.2$ and $r_{flat}=0.5$), figure~\ref{FIG:14}, we can observe the same trends compared to the DIII-D study; the growth rate of the $n=2$ AE reaches a local maximum if both beams temperatures are the same (panels a and c) increasing as the variable NBI injection intensity is enhanced (panels b and d). 

\subsection{Two fluid effects} 

\begin{figure}[h!]
\centering
\includegraphics[width=0.5\textwidth]{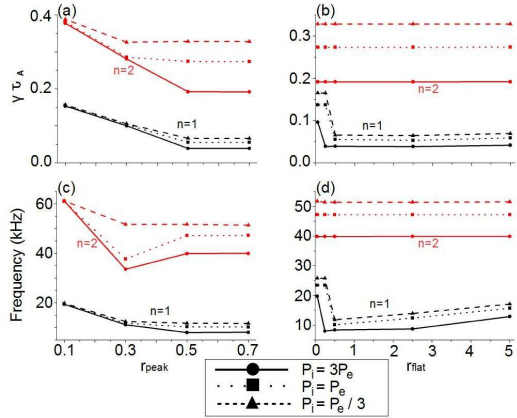}
\caption{AE growth rate (a and c) and frequency (b and d) of $n=1$ and $n=2$ AEs in the variable NBI B EP density profile study if the two fluid effects are included in the simulations. The solid lines show multiple NBI simulations with an electron pressure 3 times the proton pressure, dotted lines if the electron and ion pressure is the same and dashed lines if the electron pressure is 3 times smaller than the ion pressure.}
\label{FIG:15}
\end{figure}

Figure~\ref{FIG:15} shows $n=1$ and $2$ AEs growth rate and frequency if the two fluid effects are included in the model. The diamagnetic currents avoid the stabilization of the $n=1$ AE, stable in the previous simulations if $r_{peak} > 0.5$ or $r_{flat} < 0.5$, leading to the destabilization of a AE with a frequency around the $10$ kHz. In addition, the $n=2$ AE is further destabilized if $r_{peak} > 0.3$. It should be noted that the AE growth rate and frequency increases as the diamagnetic currents are enhanced (the pressure ratio between the thermal electrons and ions is larger).

\begin{figure}[h!]
\centering
\includegraphics[width=0.5\textwidth]{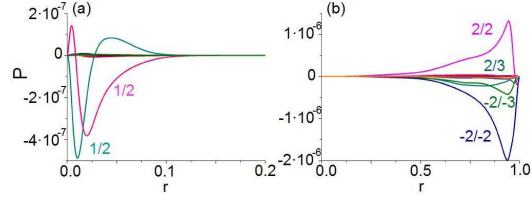}
\caption{Pressure eigenfunctions of $n=1$ (a) and $n=2$ (b) AEs in the multiple NBI simulations including the effect of the diamagnetic currents for $r_{flat} = 1$ ($P_{i} = 3 P_{e}$).}
\label{FIG:16}
\end{figure}

Figure~\ref{FIG:16} indicates the pressure eigenfunctions of $n=1$ (a) and $n=2$ (b) AEs in the multiple NBI simulations including the effect of the diamagnetic currents for $r_{flat} = 1$ ($P_{i} = 3 P_{e}$). A $1/2$ BAE/BAAE is destabilized near the magnetic axis and a $2/2-2/3$ TAE is unstable in the plasma periphery.

In summary, if the diamagnetic currents are strong enough, the stabilizing effect of the multiple beam configuration can be overcome and AEs are destabilized. Such AEs show a weaker dependency with the density profile of the EP because the main driver is the diamagnetic current, associated with the parameters of the thermal plasma, particularly the ratio between the electron/ion pressure.

\subsection{Helical couplings effect}

\begin{figure}[h!]
\centering
\includegraphics[width=0.5\textwidth]{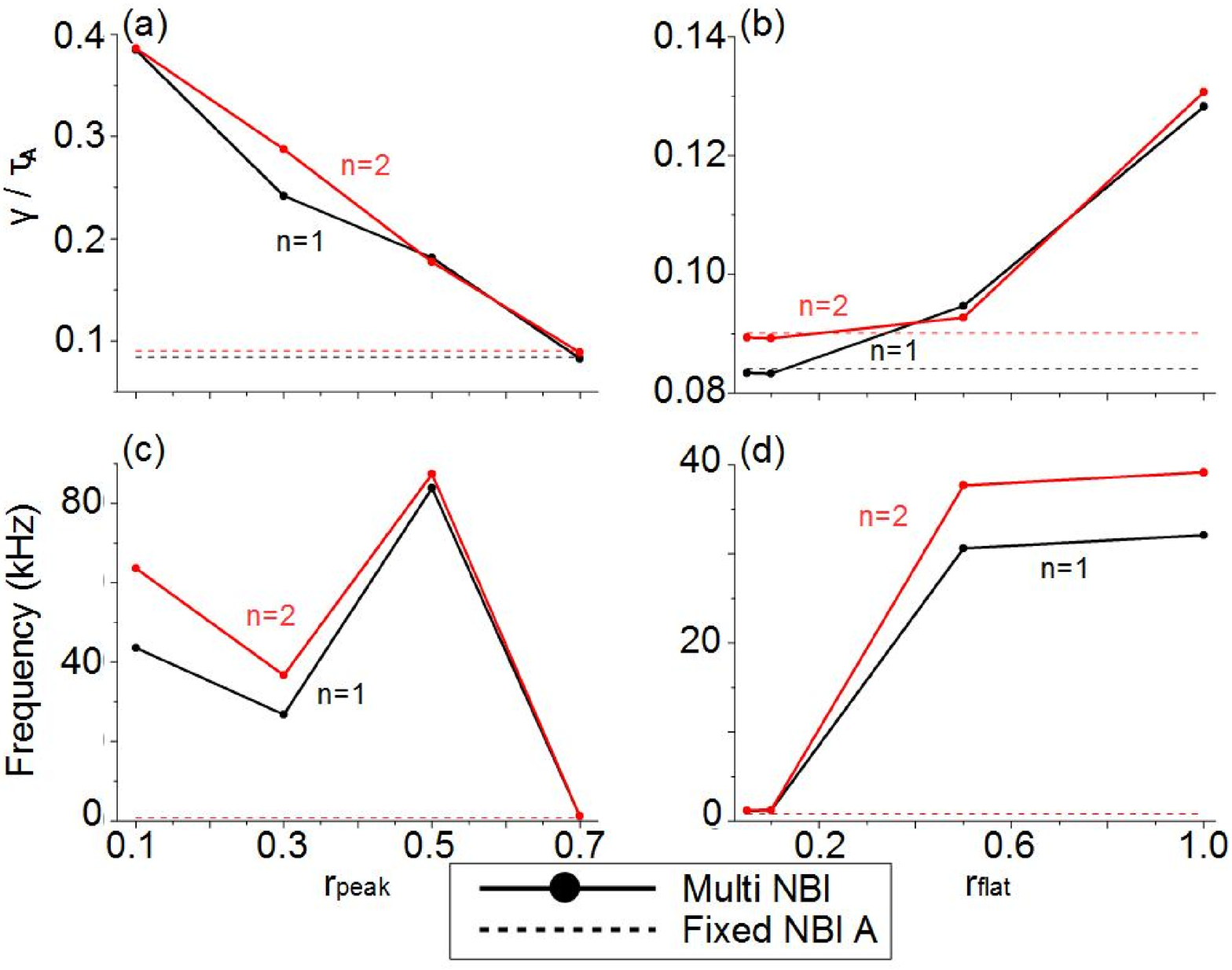}
\caption{AE growth rate (a and c) and frequency (b and d) of $n=1$ and $n=2$ AEs in the study where the density profile of the variable NBI EP is modified and the helical couplings are included in the simulations. The solid lines show the multiple NBI simulations and the dashed lines the fixed NBI A simulations.}
\label{FIG:17}
\end{figure}

Figure~\ref{FIG:17} shows the $n=1,-9,11$ and $2,-8,12$ AE growth rates and frequencies if the helical couplings are included in the model. The profiles trends are similar to the simulations without helical couplings, although the $n=1,-9,11$ AE growth rate and frequency are higher for all $r_{peak}$ and $r_{flat}$ values. On the other hand, $n=2,-8,12$ AE growth rate and frequency are smaller in all simulations. Therefore, $n=1,-9,11$ ($n=2,-8,12$) AEs are less (more) sensitive to the multiple NBI damping effect. It should be noted that the $n=2,-8,12$ AEs as well as the $n=1,-9,11$ AEs are stable if the density profile of the variable NBI EP is flat enough compared to the fixed NBI ($r_{flat} < 0.1$) so the NBI operational regime is in the damped regime.

\begin{figure}[h!]
\centering
\includegraphics[width=0.5\textwidth]{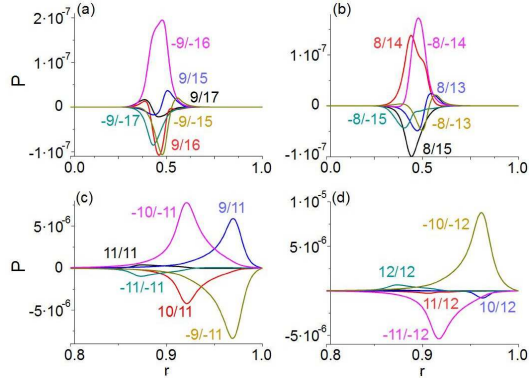}
\caption{Pressure eigenfunctions of $n=1,-9,11$ (a) and $n=2,-8,12$ (b) AEs in the multiple NBI simulations including helical couplings for $r_{peak}=0.5$. Pressure eigenfunctions of $n=1,-9,11$ (a) and $n=2,-8,12$ (b) AEs if $r_{flat}=0.1$.}
\label{FIG:18}
\end{figure}

Figure~\ref{FIG:18} shows the pressure eigenfunctions of the $n=1,-9,11$ and $2,-8,12$ AEs for $r_{peak}=0.5$ and $r_{flat}=0.5$ simulations. If $r_{peak}=0.5$, the $n=1,-9,11$ ($n=2,-8,12$) AEs are involve coupling between the modes $9/15$ and $9/16$ ($8/14$ and $8/15$) in the middle plasma region. If $r_{flat}=0.1$, in the multiple NBI damped regime, the AEs are stable and an $n=11$ ($n=12$) ballooning mode is destabilized in the plasma periphery by the coupled $9/11$ to $11/11$ modes ($10/12$ to $12/12$).

In summary, for LHD low density / magnetic field discharges with multiple NBI lines operating in a non damped regime, the simulations suggest a reinforcement of the AEs growth rate if the NBIs beam temperature is similar. To operate with multiple NBI components in the damped regime requires a density profiles of the NBIs EP with different flatness or deposition region ($r_{peak} > 0.5$ or $r_{flat} < 0.5$), stabilizing $n=1$ AE, although $n=1$ low frequency AE are destabilized if the effect of the diamagnetic currents is included in the model. In addition, if helical couplings effects are considered, the multiple NBI damping effect for the $n=1,-9,11$ helical family is weaker. The multiple NBI components operate in the interaction regime for the $n=1$ ($n=2$) modes if $r_{peak}$ is between $0.3$ and $0.5$ ($0.3$ and $0.4$) or $r_{flat}$ is between $0.5$ and $3$. On the other hand, $n=2$ AE are not stabilized so no multiple NBI damped regime is observed and $n=2$ TAE are further destabilized by the effect of the diamagnetic currents, although $n=2,-8,12$ AEs are stable if the effect of the helical couplings are added in the simulations.

\section{Conclusions and discussion \label{sec:conclusions}}

The simulations performed in the present study explore the effect of multiple energetic particle components on AE stability. The combination of different NBI lines can lead to a further destabilization of unstable AEs or stabilizing effects depending on the NBI parameters: $\beta_{f}$, beam energy or deposition profile. If the combination of the NBI lines suppresses the AE growth rates compared to the AEs destabilized by single NBI components, the NBIs operate in the ''multiple NBI damped regime''. On the other hand, if the combination of the NBI lines reduces the AE growth rates compared to the AEs destabilized by single NBI components, the NBIs operate in the ''multiple NBI interaction regime''.

We also studied the effect of the NBI components configuration on the AE growth rates in the damped and interaction regimes, identifying the most unstable combinations that should be avoided as well as the combinations that maximize the multiple NBI damping effect. In the non-damped regime the largest AE growth rate is observed if both NBI components have similar beam energy, and further enhanced if the injection intensity increases or the slope of the NBI driven EP density profiles are steeper. On the other hand, the interaction and damped regimes are associated with NBI components with different NBI driven EP density profiles, in particular, if one of the EP density profiles is flatter than the other or the NBIs are deposited in different regions of the plasma. In the interaction regime the AEs growth rate decreases if the beam energy of the NBI components is similar, the difference of flatness between NBI driven EP density profiles is larger or the $\beta_{f}$ of the NBI that drives the flatter EP density profile increases. Consequently, the role of the energetic particle resonance with normally stable Alfven waves is essential to understand the damping or enhancement of the AEs in multiple beam configurations.

DIII-D high poloidal $\beta$ discharges with multiple NBI lines operates in the interaction regime if the slope of the variable NBI driven EP density profile is weaker than $r_{flat} = 0.5$, although no full AE stabilization is observed for any combination of the NBI components (no damped regime). In the interaction regime, if both NBI energetic particle populations have similar beam energy, $T_{b} = [45,65]$ keV, or the $\beta_{f}$ of the variable NBI is enhanced, the $n>1$ AEs growth rate decreases up to a $5 \%$, except for $n=4$ AE showing a larger decrease of the growth rate and frequency caused by a transition between a $n=4$ EAE to a $n=4$ TAE.

LHD low density / magnetic field discharges with multiple NBI components operate in the damped regime if the variable NBI is deposited between the middle and outer plasma region ($r_{peak} > 0.5$) or the EP density profile slope is weak ($r_{flat} < 0.5$), leading to the stabilization of an $n=1$ AE although no stabilization is observed for $n=2$ AE. The multiple NBI components operate in the interaction regime for the $n=1$ ($n=2$) modes if $r_{peak}$ is between $0.3$ and $0.5$ ($0.3$ and $0.4$) or $r_{flat}$ is between $0.5$ and $3$. If the effect of the diamagnetic currents are included in the model, the full $n=1$ AE stabilization is not attained in the damped regime because a $1/2$ BAE/BAAE with $f \approx 10$ kHz is destabilized near the magnetic axis. In addition, a $2/2-2/3$ TAE is further destabilized in the plasma periphery. If the helical couplings are included in the simulations, the $n=1,-9,11$ helical family shows a weaker multiple NBI damping effect although it is enhanced for the $n=2,-8,12$ helical family, also stabilized.

Following up the results of the multiple NBI components study, AEs stability in tokamak and stellarators show potential optimization trends in discharges with several NBIs components if their configuration is in accordance with the requirements of a multiple NBI damped and interaction regimes. Present and future nuclear fusion devices use intense heating sources, particularly NBI, to reach the plasma temperature requirements of high $\beta$ operation leading to the destabilization of Alfven modes. Such AE activity can be minimized or even suppressed by the interaction of multiple NBI populations, although the viability of this optimization tool must be confirmed in dedicated experiments.

\begin{acknowledgments}
This material based on work is supported both by the U.S. Department of Energy, Office of Science, under Contract DE-AC05-00OR22725 with UT-Battelle, LLC and U.S. Department of Energy, Oﬃce of Science, Oﬃce of Fusion Energy Sciences, using the DIII-D National Fusion Facility, a DOE Oﬃce of Science user facility, under Award No. DE-FC02-04ER54698. DIII-D data shown in this paper can be obtained in digital format by following the links at https://fusion.gat.com/global/D3D\_DMP. This research was sponsored in part by the Ministerio of Economia y Competitividad of Spain under project no. ENE2015-68265-P, National Natural Science Foundation of China Grant No. 11575249, National Magnetic Confinement Fusion Energy Research Program of China under Contract Nos. 2015GB110005, 2015GB102000. The authors would like to thank A. Garofalo, J. Qian, C. Holcomb, A. Hyatt, J. Ferron and C. Collins for their role creating the profiles and kinetic EFIT used in the study.  \\
\end{acknowledgments}

\textbf{Disclaimer:}
This report was prepared as an account of work sponsored by an agency of the United States Government. Neither the United States Government nor any agency thereof, nor any of their employees, makes any warranty, express or implied, or assumes any legal liability or responsibility for the accuracy, completeness, or usefulness of any information, apparatus, product, or process disclosed, or represents that its use would not infringe privately owned rights. Reference herein to any specific commercial product, process, or service by trade name, trademark, manufacturer, or otherwise does not necessarily constitute or imply its endorsement, recommendation, or favoring by the United States Government or any agency thereof. The views and opinions of authors expressed herein do not necessarily state or reflect those of the United States Government or any agency thereof.

\end{document}